\documentclass[twocolumn,aps,pre,superscriptaddress,amsmath,amssymb]{revtex4-2}
\usepackage[T1]{fontenc}

\usepackage{siunitx}
\usepackage{gensymb,amssymb}
\usepackage[bottom]{footmisc}
\usepackage{lipsum}
\usepackage{graphicx}
\usepackage{dcolumn}
\usepackage{bm}
\usepackage{flushend}
\usepackage{xcolor}
\usepackage{lipsum}
\usepackage{multirow}
\usepackage{tcolorbox}
\usepackage{esvect}
\usepackage{textcomp}
\usepackage[shortlabels]{enumitem}

\usepackage[colorinlistoftodos]{todonotes}

\usepackage{hyperref}
\hypersetup{backref=true, 
  pagebackref=true
  urlcolor= blue, 
  linkcolor= blue, 
}

\begin{document}

\title{GRASPion: an Open-Source, Programmable Brainbot for Active Matter Research}

\author{F. Novkoski}
\affiliation{PULS, Institute for Theoretical Physics, FAU Erlangen-Nurnberg, 91058, Erlangen, Germany}
\affiliation{GRASP, Institute of Physics B5a, University of Li\`ege, B4000 Li\`ege, Belgium}
\author{M. M\'elard}
\affiliation{GRASP, Institute of Physics B5a, University of Li\`ege, B4000 Li\`ege, Belgium}
\author{M. Delens}
\affiliation{GRASP, Institute of Physics B5a, University of Li\`ege, B4000 Li\`ege, Belgium}
\author{F. W\'ery}
\affiliation{GRASP, Institute of Physics B5a, University of Li\`ege, B4000 Li\`ege, Belgium}
\author{M. Noirhomme}
\affiliation{GRASP, Institute of Physics B5a, University of Li\`ege, B4000 Li\`ege, Belgium}
\author{J. Pande}
\affiliation{Department of Physical and Natural Sciences, FLAME University, Pune, India}
\author{A. Maier}
\affiliation{Pattern Recognition Lab, Department of Computer Science, FAU Erlangen-Nurnberg, Erlangen, Germany}
\author{A. S. Smith}
\affiliation{PULS, Institute for Theoretical Physics, FAU Erlangen-Nurnberg, 91058, Erlangen, Germany}
\affiliation{Group for Computational Life Sciences, Division of Physical Chemistry, Ru\dj{}er Bo\v{s}kovi\'c Institute, Zagreb 10000, Croatia}
\author{N. Vandewalle}
\affiliation{GRASP, Institute of Physics B5a, University of Li\`ege, B4000 Li\`ege, Belgium}

\begin{abstract}
We present the GRASPion, a compact, open-source bristlebot designed for the controlled study of active matter systems. Built around a low-cost Arduino-compatible board and modular 3D-printed components, the GRASPion combines ease of use, programmability, and mechanical versatility. It features dual vibrating motors for self-propulsion, integrated sensors for local interaction, and customizable firmware enabling various motion modes, from ballistic to diffusive regimes. The robot is equipped with onboard IR communication, color and proximity sensors, and a magnetometer, allowing for real-time interaction and complex collective behaviors. With a runtime exceeding 90 minutes and reproducible fabrication, the GRASPion provides a robust and scalable platform for both educational and research applications in out-of-equilibrium physics. This article details the mechanical and electronic design and software architecture of the GRASPion, and illustrates its capabilities through prototypical experiments relevant to active matter.
\end{abstract}

\maketitle

\section{Introduction}

The field of active matter studies the properties and dynamics of large numbers of individual agents that draw energy from their environment and convert it into self-propulsion or interactions with others. This typically leads to out-of-equilibrium states, most notably collective motion, some common examples of which are observed in biological systems such as flocks of birds or schools of fish. The field has drawn large theoretical attention, but in recent years, controlled experimental studies have also developed, particularly in the contexts of dry granular matter, colloidal suspensions, and robotics.

Among the various experimental platforms, centimeter-scale bristlebots have emerged as a versatile and accessible model system for active matter research. Locomotion in these robots is typically achieved through a vibrating motor that transfers momentum to elastic bristles attached to the body, converting vibration into directed motion. Early implementations relied on 3D-printed asymmetric bodies placed on vibrating plates, producing chiral motion via frictional asymmetries, but required external driving from a mechanical shaker~\cite{Scholz2018}. The introduction of self-contained commercial units, such as the Hexbug\textregistered, made it possible to conduct autonomous experiments at low cost~\cite{Dauchot2019,Baconnier2022}, while custom-built variants~\cite{Deblais2018,Li2021,Boudet2021} enabled specific designs for physics-oriented studies. However, because active matter physics focuses not only on individual propulsion but also on interactions between agents, the need arose for programmable bristlebots capable of sensor-mediated interactions and reproducible behaviors. Several platforms, including the Swarmodroid \cite{Dmitriev2023} and the brainbot \cite{Noirhomme2025}, addressed some of these requirements, but limitations persisted in achieving fully controlled, directed and fast motion.

\begin{figure}[t!]
  \includegraphics[width=\columnwidth]{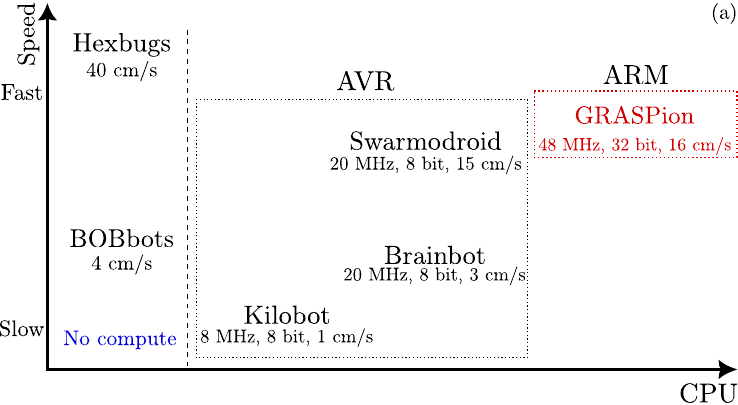}\vspace*{0.8em}
  \includegraphics[width=\columnwidth]{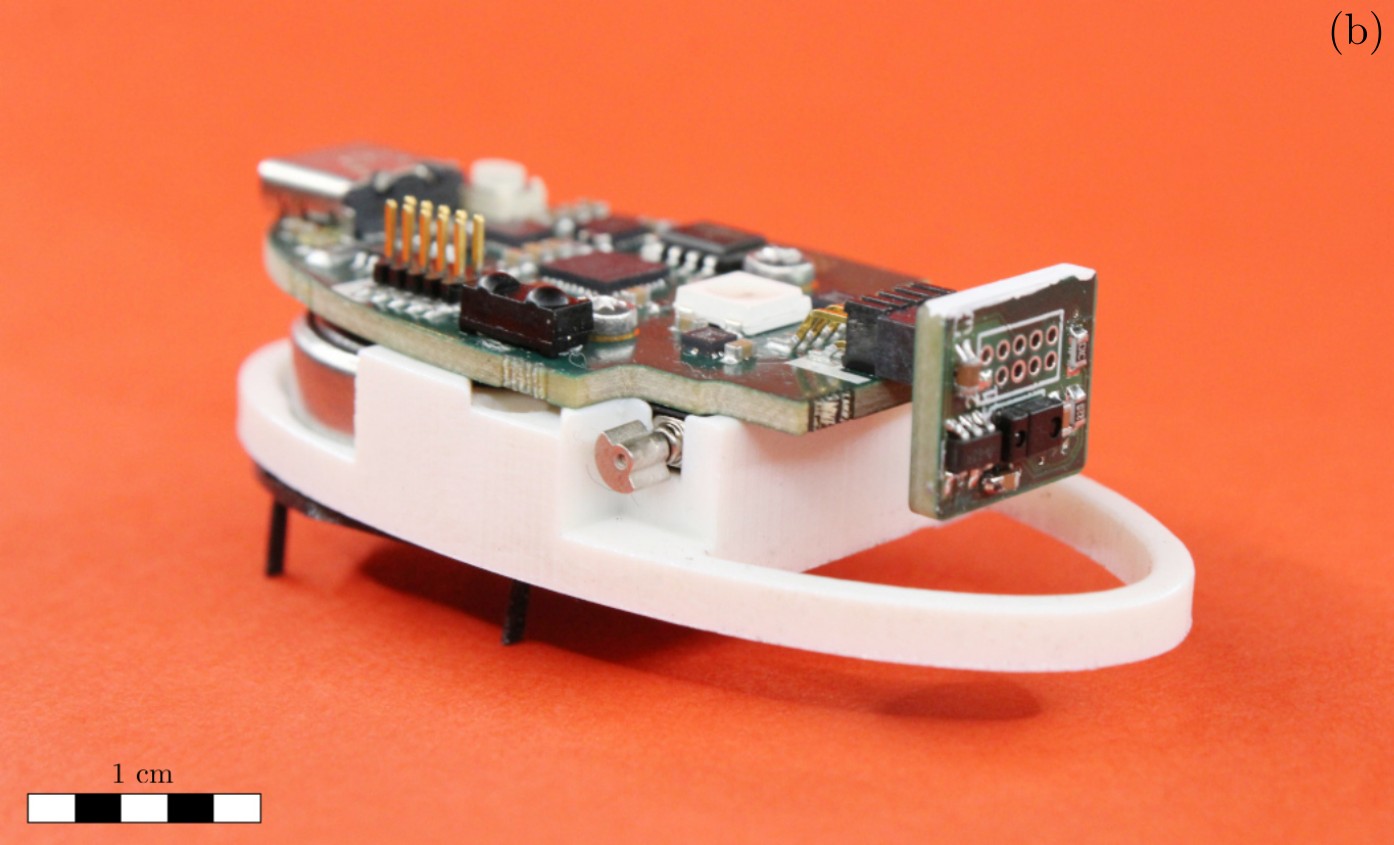}
  \caption{\label{fig:bbot}Top: The evolution of speed and processing power of different varieties of bristlebots. Bottom: Photo of the GRASPion, a fully programmable bristlebot.}
\end{figure}

Today, a broader technological need is emerging in the field of active matter: experimental capabilities must keep pace with increasingly sophisticated theoretical and numerical models. Figure~\ref{fig:bbot}a highlights a trend among existing bristlebot platforms, showing a positive correlation between achievable speed and available onboard computational power. This reflects a growing demand for faster and more intelligent active particles, capable of extensive data collection, high responsiveness, and adaptive behaviors, including onboard machine learning and real-time decision-making. Such advances would open new avenues for studying complex collective dynamics and emergent phenomena at the interface of physics, robotics, and computation.

In this paper, we present an easy-to-use, Arduino-based bristlebot that is both low-cost and compact, with easily replaceable 3D-printed parts and open-source firmware. The robot, termed the GRASPion, can be remotely controlled, can execute preprogrammed motion (including random diffusive behavior), and is easily rechargeable, with a runtime exceeding 90 minutes. It is designed for rapid deployment in both teaching and research, enabling controlled experiments in active matter while supporting flexible extensions via its modular mechanical, electronic, and software architecture. Unlike other open-source bristlebots, the GRASPion is also commercially available as a ready-to-use product, lowering the barrier to entry for laboratories and educators interested in exploring the physics of active systems.

\begin{figure}[t!]
  \includegraphics[width=\columnwidth]{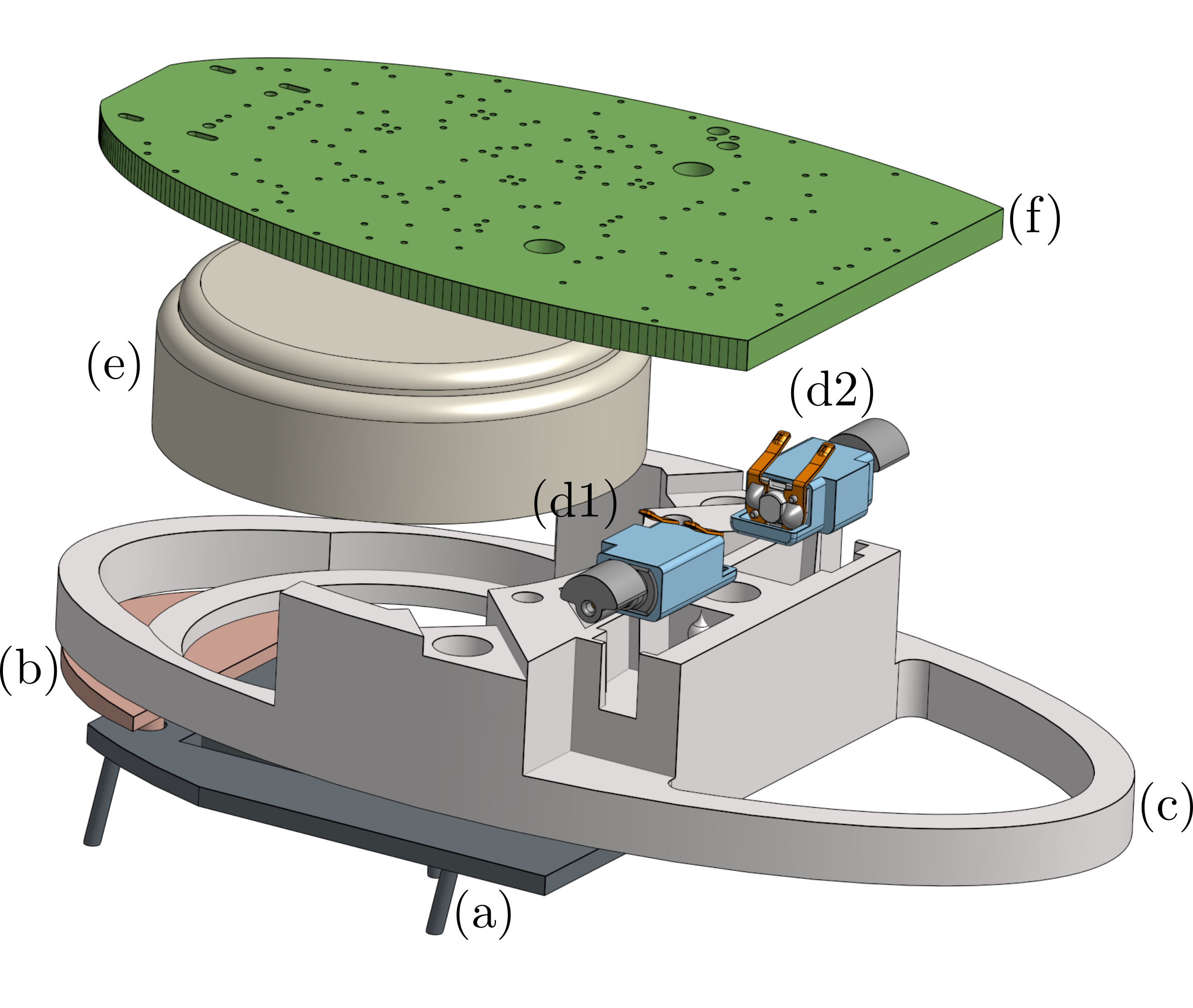}
  \caption{\label{fig:exploded}Exploded view of all bot parts: (a) the PLA leg plate, (b) wedge plate,
    (c) the ABS printed body, (d) the two vibrating motors with spring contacts, (e) the lithium ion battery, (f) the bot circuit board.}
\end{figure}

In the following sections, we will describe the main components of the bot, including a technical description of the mechanical and electronic components, as well as provide several use cases of the bots, with the accompanying code.

\section{Bot description}

A representative image of the GRASPion is presented in Fig.~\ref{fig:bbot}b, with its key components highlighted. The robot features an elliptical footprint, with overall dimensions of $60 \times 30 \times 15$ mm$^{3}$ and a total mass of 17 g, including all components. The chassis (white in the figure) serves as the main structural element, onto which the electronic circuit board is mounted, and the legs are attached underneath. The legs are a critical functional component, as they enable self-propulsion through frictional interaction with the substrate. Their geometry and material properties play a central role in determining the efficiency and reliability of the robot’s locomotion. The only electromechanical elements of the GRASPion are two vibration motors, embedded directly within the chassis, which serve as the actuation mechanism driving the motion.

The intelligence of the GRASPion stems from its onboard electronic circuit. The uploaded software enables full control over the robot’s motion by independently addressing each motor via dedicated drivers. Additionally, the microcontroller processes data from the onboard sensors, allowing the bot to perform real-time decision-making based on environmental input. The entire system is powered by a rechargeable lithium-ion battery, ensuring autonomous operation over extended durations.

With a general overview of the bot given, we will first focus on the mechanical parts in Section~\ref{subsec:body}, followed by a more in-depth discussion of the circuit in Section~\ref{subsec:circuit}, and finally concluding with the final layer, the bot firmware in Section~\ref{subsec:soft}.

\subsection{Mechanical parts}\label{subsec:body}

The body of the bot is of an ellipsoidal shape with major and minor axes of length $60$ and $30$ mm, respectively, with the total height of the body being $9$ mm, without the legs. The body is 3D printed out of ABS plastic, with the STL files and print settings for all parts available through a public repository~\cite{GithubGRASP}. The body can be freely modified to accommodate experimental requirements, such as to enable physical connections between bots, break body symmetry, add chirality, or transform into a purely circular shape (corresponding circular circuits can be made available upon request). The complete view of all parts of the bot is shown in Figure~\ref{fig:exploded}. The body has a total mass of $6$ g, and has been optimized to reduce weight as much as possible while retaining structural integrity. The only required post-printing step is tapping the holes that hold the circuit and legs to the body.

In order to induce motion, the bot relies on four legs, printed out of PLA plastic, with a diameter of $0.8$ mm, and a $12\degree$ angle with respect to the vertical. The leg angle has a strong impact on
the locomotion of the bot, and can be varied in order to serve different purposes. The legs are
printed as part of a leg-plate (part a in Fig.~\ref{fig:exploded}) which attaches to the body of the bot
through two M2 screws, making them easily replaceable in case of breaking. We have found that it is
crucial that the leg plates are printed out one at a time (not multiple legs in parallel layer by
layer) in order to avoid weak layer adhesion on the legs. Between the leg-plate and the body, a wedge (part b in Fig.~\ref{fig:exploded}) is added in
order to provide a forward pitch to the bot, allowing for a more directed and controlled forward
motion. The parameters of the legs have been carefully chosen in order to achieve reproducible and
well-controlled bot trajectories. In addition to cylinder legs, a rectangular form is also provided, which enhances the durability of the legs as well as bot speed, at a slight cost of trajectory
control.

As mentioned above, the bot is driven by two vibrating motors
(\href{https://www.vybronics.com/erm-cylindrical-vibration-motors/spring-contacts/v-z3th8b171700l}{VZ3TH8B171700L}),
which are enclosed in two corresponding compartments in the body of the bot, observed in
Fig.~\ref{fig:exploded}. The motors use spring contacts to receive power, meaning that they are
detachable from the circuit, allowing for easy replacement. The applied voltage and its polarity to each motor, and
thus the speed and rotation direction, can be controlled directly through the firmware (see below), with the motors being
independent from each other.

\begin{figure}[t!]
  \includegraphics[width=0.9\columnwidth]{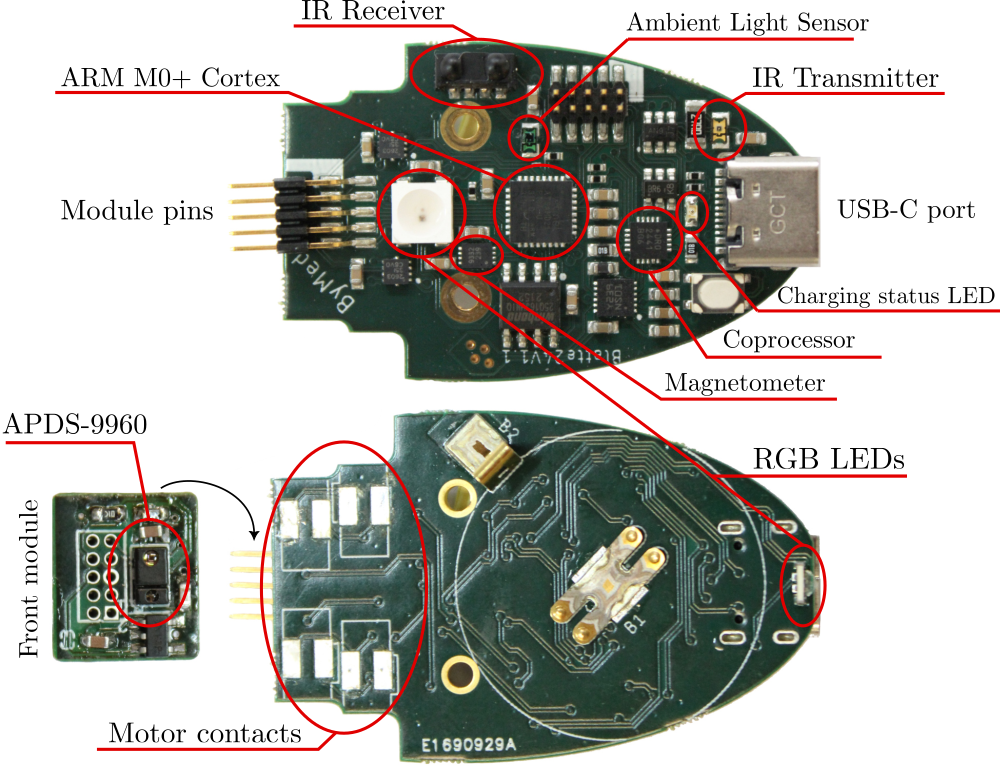}
  \caption{\label{fig:circuit}Top and bottom of the GRASPion circuit board. On the top circuit, we find most of the components, including the M0+ Cortex processor, an RGB LED, and the USB-C socket. On the bottom are located the contacts for the motors, which are connected through their spring contacts once the board is mounted. Additionally, we find the battery contact, along with an additional rear RGB LED. On the other end, we find the pins for the detachable front module, which contains the APDS-9960 color and proximity sensor.}
\end{figure}

\subsection{Electronics}\label{subsec:circuit}
The control over the bot's motion, as well as all its other capabilities, comes from the onboard circuit. The onboard circuit is based around an Adafruit QtPy SAMD21 clone, which relies on the ARM Cortex M0+ processor. A custom firmware is flashed a single time onto a low-power AVR coprocessor (AVR32DD20), which handles all low-level processes, including charging, IR reception, and transmission. This allows a higher-level programming of the bot through the standard Arduino IDE and the accompanying libraries, using the onboard USB-C port. The GRASPion is thus currently the only Arduino-based bristlebot, making its deployment and control fast and easy.

The circuit also comes equipped with:
\begin{itemize}
\item an IR receiver and transmitter,
\item a 3-axis magnetometer,
\item two Neopixel RGB LEDs, one on top and one at the back,
\item a Flash memory ($2$M-byte),
\item an ambient light sensor mounted on top, and
\item a front module that can be exchanged and which currently contains:
  \begin{itemize}
  \item a proximity sensor,
  \item a color detector (RGB), and
  \item a gesture sensor (left, right, up, down).
  \end{itemize}
\end{itemize}
The last of the listed components, the module, is a small add-on circuit board, as seen in Figure~\ref{fig:circuit}. While it currently contains the above-mentioned detectors, this can be customized and exchanged for other types of sensors, such as an accelerometer, for example. This add-on interface also serves as the contact point for a Bluetooth module that is currently under testing, and which will allow for a more robust communication protocol between the bots as compared to IR. The complete pinout and design of the circuit are available on the public repository~\cite{GithubGRASP}.

\subsection{Software}\label{subsec:soft}
As mentioned previously, the bot is completely controlled through standard Arduino code, meaning it
requires only knowledge of C++. A complete guide on which libraries are needed for the programming
of the board, as well as which configuration should be used in the Arduino IDE in order to do it, is given in
the repository. An additional advantage is that, through the use of a USB charging hub, it is possible to bulk flash large numbers of bots at once, with the code needed for this being made available as well.

Within the main loop of the code, access to all of the sensor data is available, as well as direct control over the two motors. This makes it extremely simple to direct the bot with a pre-programmed motion, make it react to its environment, or simply remote control it. It is also possible to induce random motion (through the use of the random C function) -- however, for this a seed has to be given, which is done through the use of the analog pin on the IR sensor. Below, several examples are given for the bot motion, with references to the accompanying code.

\section{Examples of use cases}

In this section, we present three simple use cases of bot motion. All of the code can be found in the example directory of the repository.

\begin{figure}[t!]
  \includegraphics[width=\columnwidth]{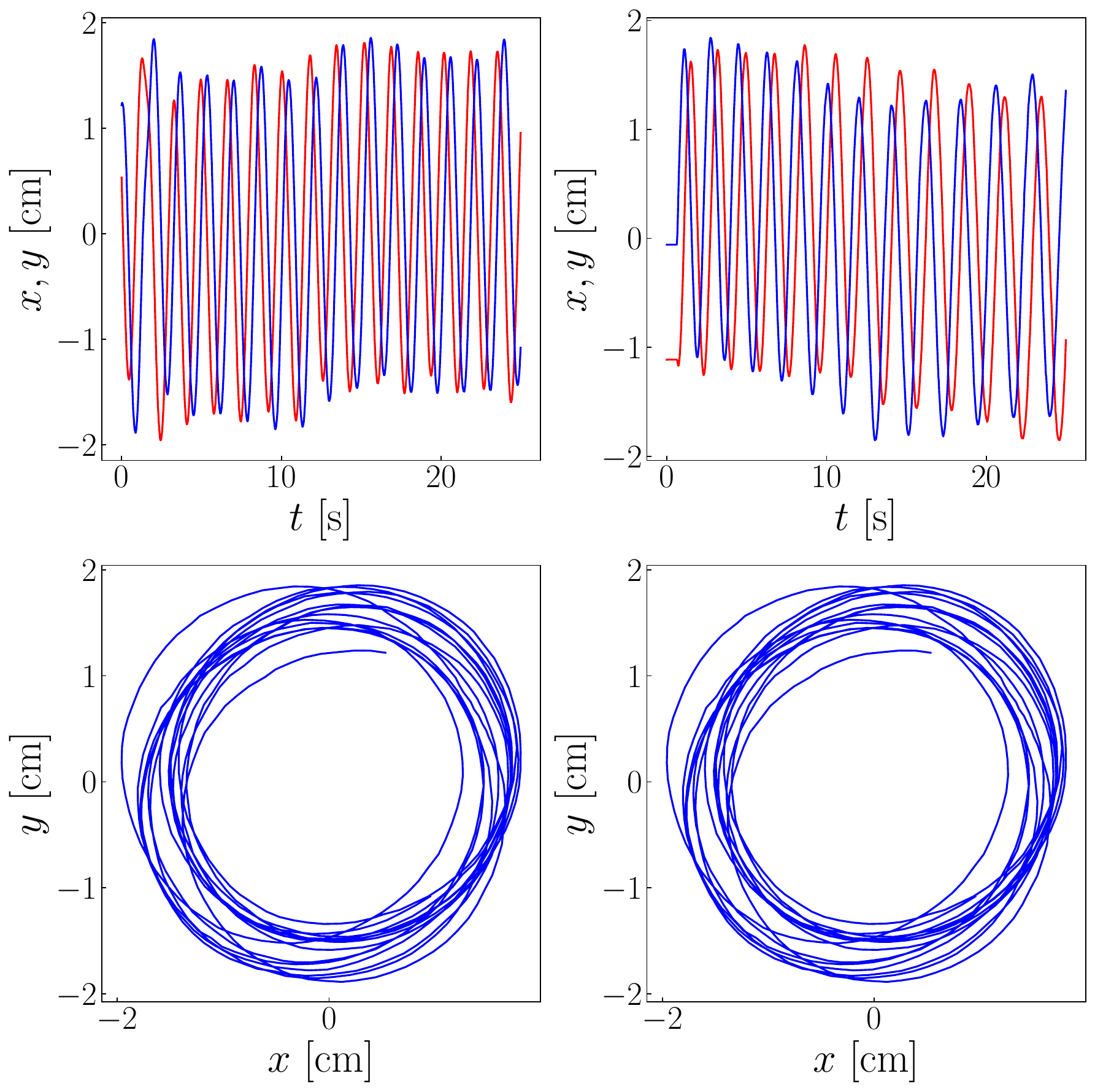}
  \caption{\label{fig:turning}Top: the $x$ and $y$ coordinates of the bot center of mass as it turns
    clockwise (top left) and counter-clockwise (top right). Bottom: the trajectory in the $xy$
    plane for clockwise (left) and counter-clockwise (right) motion.}
\end{figure}

\subsection{Simple remote control}
Using the IR receiver on the bot board, we are able to remotely control it using a standard remote, as
is used for television sets, for example. The remote control emits an IR signal, which the bot decodes
into hexadecimal values and can then easily be handled in the Arduino code.

The essential example of this is remotely controlling the motion of all the bots at the same time. In order to turn the bot, we need to power only one of the motors (turning on the left motor makes the bot turn right, and vice versa). In the given code, the function \texttt{buzCw} takes two arguments, the first one corresponding to a power level, with allowed values from $0$ to $127$, and the second to the pin of the
motor we wish to turn on, either \texttt{BUZL\char`_PIN} for the left, or \texttt{BUZR\char`_PIN} for the right motor, which are defined at the beginning of the code base on the pinout. For completeness, a secondary function \texttt{buzCcw} is also present, turning the motor in the counter-clockwise direction; however, this has the same physical consequence as \texttt{buzCw}.

Thus, to turn, we first have to call the function \texttt{buzStop(0)} to turn off all motors, and then simply call \texttt{buzCw} with the appropriate arguments. Already, this type of basic motion allows for accessible studies of chiral motion and chiral active particles, providing immediate experimental relevance, as well as opening up the possibility of controlled and programmable chiral particles. 

In Figure~\ref{fig:turning}, we plot the measured $x$ and $y$ coordinates of the bot's center of mass both for the left and right turns for multiple periods of rotation. Essentially, if only one motor is constantly on, the bot will simply rotate around the corresponding front leg. As we can see, this rotation is consistent over multiple periods, with very slight drift($\sim 5$ mm), and potentially
allows for studies of chiral particles. 

\begin{figure}[t!]
  \includegraphics[width=\columnwidth]{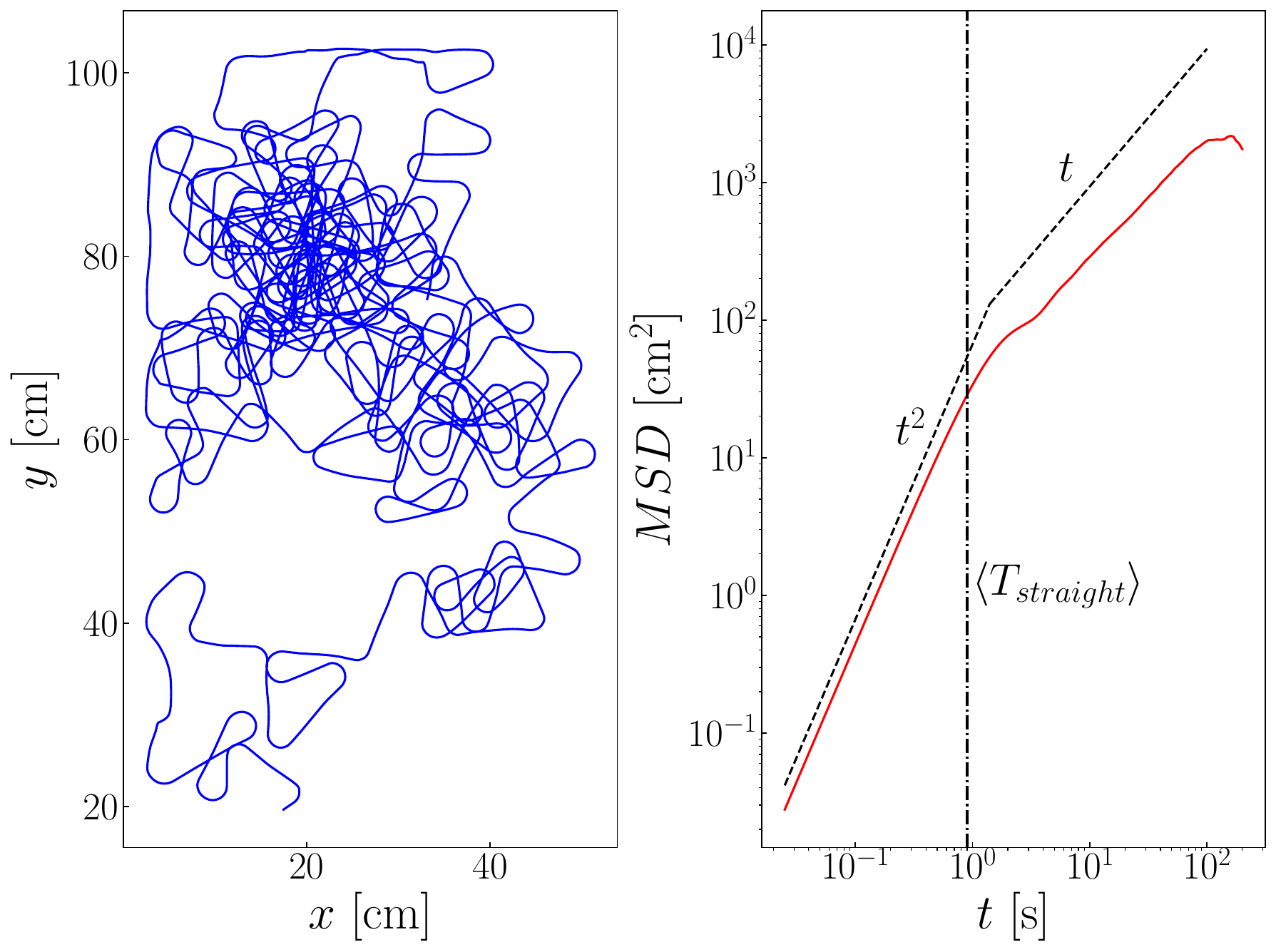}
  \caption{\label{fig:diffusion}Left: sample of the trajectory of a bot running the diffusive
    algorithm. Right: The mean square displacement of the full trajectory, with a clear transition
    from the ballistic to the diffusive regime.}
\end{figure}

\subsection{Diffusive motion}
Using the above functions, we can easily implement diffusive motion. The algorithm for this is to move straight for a time $T_\text{straight}$, which is chosen at random uniformly in the interval of $[400,1200]$ ms, after which the bot turns either left or right with a $50\%$ chance of either, and it once again turns for a random amount of time $T_\text{turn}$ chosen in the same interval as $T_\text{straight}$. It is important for the random motion that we seed it properly. This is done through the use of the native Arduino \texttt{randomSeed} function, on the pin \texttt{A0}, which corresponds to the light sensor. In order to achieve straight motion, we simply call \texttt{buzCw} for both driver pins to turn both motors on.

The above algorithm has been implemented in the \texttt{launchRT} function provided in the example code, and is activated remotely through one of the buttons on the control. By recording and tracking the bot, we can recover its trajectory, a part of which is shown in the left part of Figure~\ref{fig:diffusion}. While it is difficult to see from the trajectory itself that this motion is diffusive, it becomes much clearer once we consider the mean square displacement (MSD) in the right part of Fig.~\ref{fig:diffusion}. Here we see that at short time scales the motion is ballistic and the MSD grows with $\sim t^2$, while at longer scales, after a transition, it grows with $t$ as expected for classical diffusion.

\begin{figure}[t!]
  \includegraphics[width=\columnwidth]{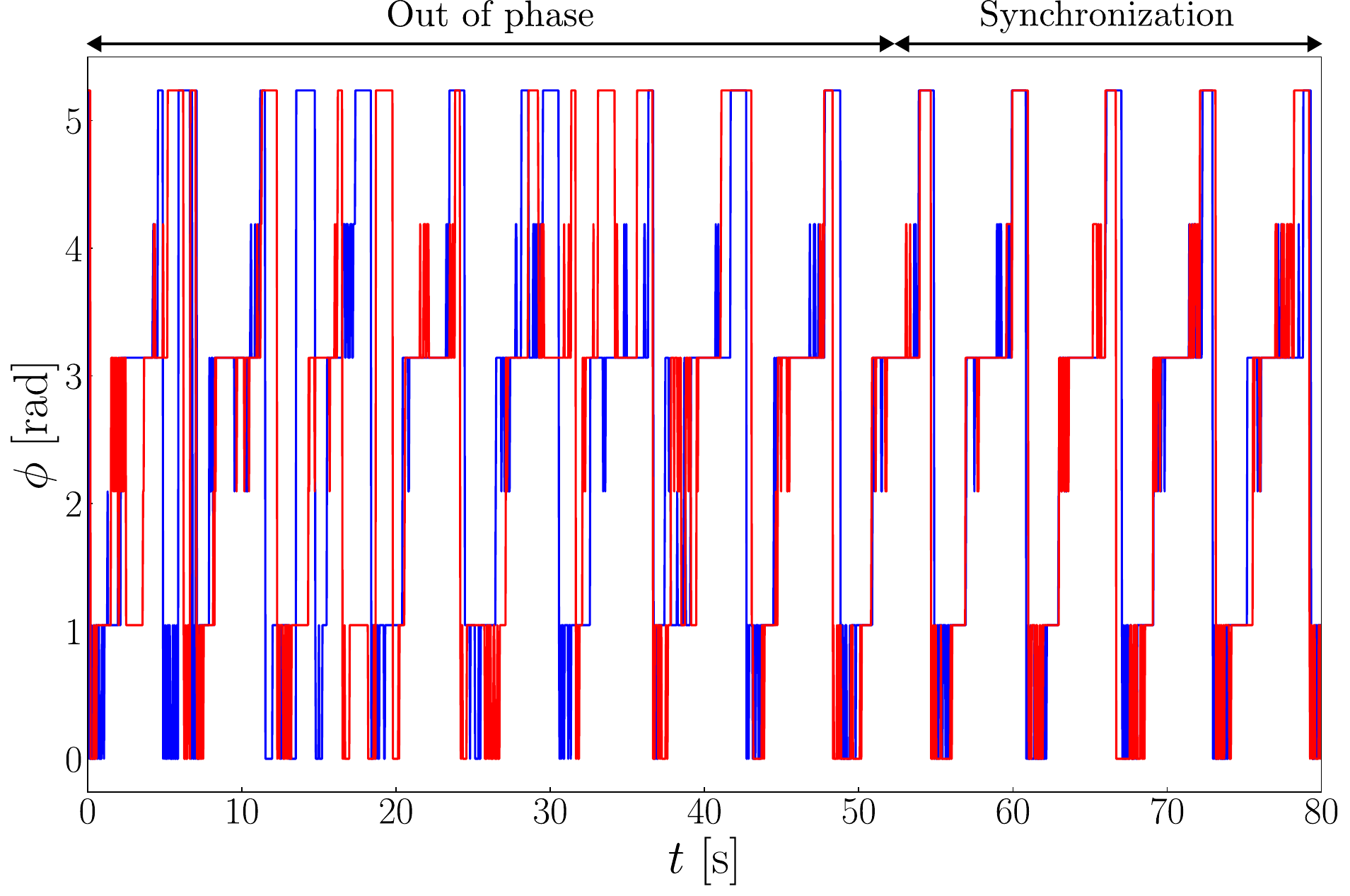}
  \caption{\label{fig:kuramoto}The phases of two bots with $6$ discrete states interacting through the prescribed discrete Kuramoto procedure. While they are out of sync initially, we see that the two oscillators synchronize both in phase and in frequency.}
\end{figure}

\subsection{Color synchronization}
Beyond self-propulsion, interaction plays a crucial role in the study of active matter. One of the emergent effects that come out of interactions of many individual agents is synchronization, as exemplified by the Kuramoto model of interacting phase oscillators~\cite{Kuramoto1975,Pikovsky2001}. Through a sufficiently strong interaction, individual oscillators can synchronize, resulting in a global oscillating frequency. When this oscillatory behavior is coupled with self-propulsion, it leads to so-called swarmalators, which can result in a wide variety of moving patterns~\cite{OKeeffe2017}.

Inspired by this, we have implemented a phase oscillatory behavior and colour detection in the bot, with the example code found in the public repository. The onboard RGB LEDs are made to go through a given cycle of discrete colours, $6$ of them in this example, which can be made to correspond to a given phase $\phi$ in the $[0,2\pi)$ interval. This colour is emitted on both LEDs, and switched with a frequency $\omega_t$. This frequency is made to vary based on the input coming from the color sensor of the front module, in order to set a
basis for synchronization. The frequency changes actively as
\begin{align}
    \omega_{t+1}=\omega_t+K\sin(\phi-\phi_d)
\end{align}
where $\phi_d$ is the detected colour coming from the outside, which is first matched with the closest one in the given list of discrete colours (done by measuring the Euclidean colour difference), and $K$ is a coupling constant that can be adjusted. Here, we rely on the \textit{Adafruit\_NeoPixel\_ZeroDMA} in order to use DDS (Direct Digital Synthesis) to have an arbitrary frequency resolution.

As an example, we show the synchronization of two stationary bots with this property, placed head to tail in order to be able to observe each other's LEDs. The front bot is made to oscillate with a fixed frequency (see the red line in Figure~\ref{fig:kuramoto}), and we can observe that the back bot synchronizes with the other one both in frequency and in phase (blue line).

\section{Conclusion}

The GRASPion offers a rare combination of mechanical simplicity, high programmability, and robust performance. From basic locomotion experiments to the emergence of complex collective behaviors, it provides a reliable and modular platform for probing the full range of active matter phenomena under controlled and reproducible conditions.

The flexibility of the GRASPion opens up unique opportunities to investigate and test frontiers in collective robotics and the emerging field of smart active matter~\cite{Bertin2025}, allowing for a model system of adaptive swarm control~\cite{Ziepke2025} and distributed learning in robotic ensembles, as well as an experimental basis for the study of classical concepts such as thermodynamic engines in the context of active matter~\cite{Pietzonka2019}, swarmalators~\cite{OKeeffe2017} and predator-prey dynamics. Moreover, thanks to its open hardware and software design, virtually any interaction law, from classical alignment and repulsion~\cite{Vicsek1995} to highly unconventional non-reciprocal couplings~\cite{Loos2020}, can be implemented, thus using programmable interactions to emulate exotic couplings rarely accessible in physical experiments. The versatility, expandability, and accessibility of this bot thus give a range of possibilities for the emulation of animate and active matter, as well as the creation of novel metamaterials, intelligent swarms, and metamachines~\cite{Volpe2025}.

The GRASPion thus stands as a versatile experimental bridge between statistical physics, robotics, and computational intelligence, paving the way for a new generation of laboratory studies on emergent behavior.

\section*{Acknowledgements}

This work is financially supported by the University of Li\`ege through the CESAM Research Unit. N.V. thanks the Fondation Francqui for support. F.N. thanks the Alexander von Humboldt Foundation for a postdoctoral fellowship.

\bibliographystyle{apsrev4-1} \bibliography{biblio}

\end{document}